\documentclass[jcp,reprint,floatfix,aip]{revtex4-1}

\usepackage{bm}
\usepackage{float}
\usepackage{graphicx}
\usepackage{color}
\usepackage{dcolumn} 
\definecolor{cfb}{rgb}{0.0,0.9,0.9}
\definecolor{orange}{rgb}{1.0,0.45,0.0}
\definecolor{lila}{rgb}{0.7,0.0,1.0}

 \usepackage{color}
 \usepackage[normalem]{ulem}
 \usepackage{cancel}
\begin{document}

\title{Self-Consistent Embedding of Density-Matrix Renormalization Group Wavefunctions in a Density
Functional Environment}

\author{Thomas Dresselhaus}
\affiliation{Westf\"alische Wilhelms-Universit\"at M\"unster,
Theoretische Organische Chemie, Or\-ga\-nisch-Chemisches Institut and Center for Multiscale Theory and 
Computation,
Corrensstra{\ss}e 40, 48149 M\"unster, Germany}

\author{Johannes Neugebauer}
\email{j.neugebauer@uni-muenster.de}
\affiliation{Westf\"alische Wilhelms-Universit\"at M\"unster,
Theoretische Organische Chemie, Or\-ga\-nisch-Chemisches Institut and Center for Multiscale Theory and 
Computation,
Corrensstra{\ss}e 40, 48149 M\"unster, Germany}

\author{Stefan Knecht}
\email{stefan.knecht@phys.chem.ethz.ch}
\affiliation{ETH Z\"urich, 
Laboratorium f\"ur Physikalische Chemie, 
Vladimir-Prelog-Weg 2, 8093 Z\"urich, Switzerland}

\author{Sebastian Keller}
\affiliation{ETH Z\"urich, 
Laboratorium f\"ur Physikalische Chemie, 
Vladimir-Prelog-Weg 2, 8093 Z\"urich, Switzerland}

\author{Yingjin Ma}
\affiliation{ETH Z\"urich, 
Laboratorium f\"ur Physikalische Chemie, 
Vladimir-Prelog-Weg 2, 8093 Z\"urich, Switzerland}

\author{Markus Reiher}
\email{markus.reiher@phys.chem.ethz.ch}
\affiliation{ETH Z\"urich, 
Laboratorium f\"ur Physikalische Chemie, 
Vladimir-Prelog-Weg 2, 8093 Z\"urich, Switzerland}

\begin{abstract}
We present the first implementation of a density matrix renormalization group algorithm embedded in 
an environment described by density functional theory. The frozen density embedding scheme is used 
with a freeze-and-thaw strategy for a self-consistent polarization of the orbital-optimized wavefunction and 
the environmental densities with respect to each other.
\end{abstract}

\maketitle

\begin{figure}[!]
  \includegraphics[width=0.6\columnwidth]{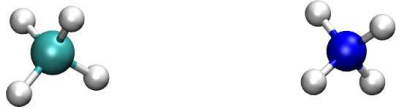}\vspace{-10pt}
  \caption{A methane molecule (left) in the presence of an ammonium ion (right) at the equilibrium 
    distance of 5.68 \AA.}
  \label{fig:methane_ammonium}
\end{figure}

\begin{figure}[!]
  \begin{minipage}{0.15\columnwidth}
  \end{minipage}
  \begin{minipage}{0.33\columnwidth}
   \centering A
  \end{minipage}
  \begin{minipage}{0.33\columnwidth}
   \centering B
  \end{minipage}
  \begin{minipage}{0.15\columnwidth}
  \end{minipage}\\
  \includegraphics[width=0.7\columnwidth]{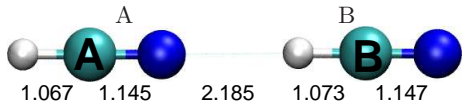}\vspace{-10pt}
  \caption{The supermolecular HCN dimer consisting of two HCN molecules denoted A and B (distances are in \AA).} 
  \label{fig:HCN_dimer}
\end{figure}

\begin{figure}[!]
  \includegraphics[width=0.8\columnwidth]{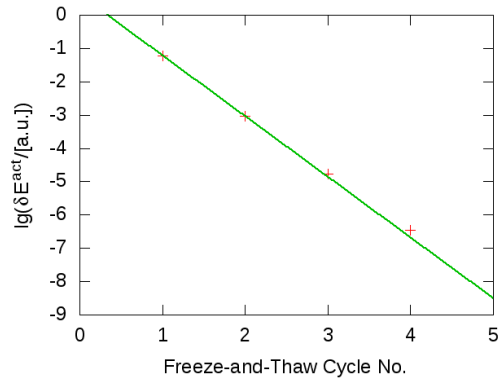}
  \caption{Error of the total electronic energy in Hartree with respect to a converged f\&t procedure for a 
    DMRG(10,9)-SCF-in-DFT calculation on a HCN molecule embedded into the environment of another HCN molecule as a 
    function of the f\&t cycle number.}
  \label{fig:FaT_convergence}
\end{figure}

Chemical reactions are local phenomena, in which usually only one or two bonds are formed or broken at a time.
For this reason, the cluster approach was established in computational chemistry, in which the atomistic rearrangements under
study are embedded in a structural model that is as large as necessary to comprise all electronic 
effects on the reactive subsystem (see, for instance, the extensive work on such cluster models for active sites of metalloproteins by
Siegbahn, Himo, and co-workers \cite{siegbahn2011}). In general, long-range electrostatic interactions may not be neglected and 
schemes have been devised to incorporate them (most prominent are quantum-mechanical/molecular-mechanical embedding \cite{senn2009} 
and polarizable continuum models \cite{Mennucci2012}). For the appropriate description of extensive electronic coupling of a subsystem to its environment
(mediated, for example, through $\pi$-conjugated molecular substructures),
more involved embedding schemes are needed.\\
Clearly, the embedding of a quantum system of interest into a more or less electronically noninnocent environment 
is the realm of open-system quantum theory \cite{aman11}. The central quantity of this well-known theory is
the reduced density matrix of a subsystem rather than the wave function of a stationary state.
This formalism has recently been adapted for an embedding framework.\cite{kniz2012,kniz2013} However, it
is not straightforward to employ (for instance, the environment had to be approximated
in a mean-field fashion). Practical suggestions for the applicability to large chemical systems have been
made.\cite{sun2014}\\
Considering the success of density functional theory (DFT) for routine applications in computational chemistry,
a feasible approach that is able to describe chemically complex and structured environments should also be based on DFT.
In the frozen density embedding\cite{weso1993} (FDE) scheme, which is based on subsystem 
DFT,\cite{sena1986,cort1991} a subsystem is calculated in a Kohn--Sham (KS) approach by adding 
an effective potential to the KS potential of the isolated subsystem. It takes into account all 
interactions with the environment, assuming a constant (frozen) environmental electron density. In 
FDE not only the exchange--correlation energy contribution, but also the non-interacting kinetic 
energy term is calculated from an approximate explicit density functional. 
No knowledge of orbitals of the supersystem is needed for an FDE/subsystem DFT calculation,
nor do the orbitals of the active subsystem need to be orthogonal to 
those of the environmental subsystems. In this way large environments can be considered.\\
Due to the limited accuracy of kinetic energy functionals for chemical systems, FDE cannot be
directly applied to subsystems which are connected via bonds with covalent character\cite{fux2008}. Extensions to 
resolve this limitation are available. Often, potential-reconstruction methods have been proposed in 
this context.\cite{ronc2008, good2010, huan2011, jaco2008, fux2010} A related embedding scheme by 
Manby \emph{at al.}\cite{manb2012} resolves this issue by imposing orthogonality constraints 
between orbitals of the active subsystem and the environment. This strategy avoids the need for 
explicit density functionals for the evaluation of the kinetic energy. 
KS orbitals, which can be conveniently obtained from 
comparatively cheap DFT calculations, are employed here.
Furthermore, a straightforward basis-set 
truncation scheme, needed for a reduction of the computational cost for the subsequent wavefunction 
calculation, is readily available.\cite{barn2013} 
In FDE, all electron-correlation effects between different subsystems are 
captured by the density functional. In principle, all correlation effects, including the polarization of the wavefunction of the active subsystem, 
can be accurately described for the (admittedly unknown) exact functional.\\
Considering the fact that at times approximate density functionals can yield rather inaccurate results,
it is desirable to describe an (active) subsystem by systematically improvable wavefunction theory (WFT).
Although based on DFT, the FDE framework also allows for treating one subsystem with a correlated 
wavefunction method and the rest of the system with DFT \cite{weso2008}. Several research groups implemented such a 
WFT-in-DFT embedding; for instance, 
to facilitate complete-active-space self-consistent field (CASSCF) 
in periodic\cite{klue2001, huan2006} and
non-periodic\cite{dada2013} DFT, M{\o}ller--Plesset perturbation theory in DFT \cite{govi1998},
coupled cluster in DFT\cite{gome2008}, CASSCF with second-order perturbation theory (CASPT2) in DFT 
\cite{kana2012,dada2013},
and most recently quantum Monte Carlo in DFT\cite{dada2014}
calculations (for an overview see Refs.\ \citenum{jaco2014,libi2014,gome2012}).\\
Within a comparatively short time, the density matrix renormalization group (DMRG) alorithm\cite{whit1992,whit1993} has become 
a standard WFT approach in quantum chemistry (for reviews see Refs.\ \citenum{ors_springer-B,mart2010,sharma2012,chan2012dmrgrev,kurashige2014,wouters2014rev}), which is
under continuous development (see Refs.\ \citenum{Wouters20141501,knec2014} for most recent work) providing interesting new perspectives 
for the description of electronic structures dominated by static electron 
correlation.\cite{spindensity2012,entanglement_letter,entanglement_bonding_2013,kura2013,harris2014,lan2014,Sharma2014,tecmer2014,wout2014}
DMRG is able to iteratively converge to the exact solution of the electronic Schr\"odinger equation in a given active orbital space with 
polynomial rather than factorial cost by which the full configuration (and thus the CASSCF) approach is plagued. However, the benefits of this iterative protocol come with a caveat that has long
prohibited the extension of its applicability to compact and strongly correlated molecules like transition
metal complexes and clusters. This caveat is rooted in the matrix-product-state structure of the optimized DMRG wave function.
As a consequence, DMRG should only be applied to pseudo-one-dimensional molecular structures.
It could be shown\cite{marti2008} that its benefits outweigh these formal objections; still,
care must be taken to control the parameters that determine the accuracy of an iterative DMRG calculation \cite{keller2014}.
It has now been fully recognized that accuracy- and performance-wise, DMRG can be a substitute for the CASSCF (and CASPT2) 
methodology pioneered by the Lund group \cite{roos08} for cases, which require much larger active orbital spaces than
those accessible by this standard approach. With DMRG, active orbital spaces are in reach that are about five times larger than those accessible 
to CASSCF. This feature makes DMRG an ideal target for a multireference WFT-in-DFT embedding method.\\
Here, we report the first implementation of a DMRG-in-DFT embedding (including orbital relaxation in the DMRG step; DMRG-SCF), 
which follows the FDE framework in a freeze-and-thaw\cite{weso1996} (f\&t) manner, to establish a fully self-consistent embedding. 
Often (but not always \cite{dada2014}) is the environment 
polarized by employing a DFT-in-DFT approach, then kept fixed in the WFT calculations, and a fixed embedding 
potential is applied\cite{gome2008,dada2013}. 
Alternatively, Carter and co-workers imposed constraints on the total\cite{govi1999} or 
environmental\cite{huan2006} density.\\
In our implementation an active subsystem (act) calculated with a 
WFT method is embedded into an environment (env) described with DFT and the following steps are carried out:\\
1) Perform DFT calculations on the isolated subsystems to
 calculate initial densities ${}^{(0)}\rho^{\text{env}}$ and ${}^{(0)}\rho^{\text{act}}$\\
2) Calculate an embedding potential for the environmental subsystem 
         ${}^{(n)}v_{emb}^{\text{env}}[{}^{(n-1)}\rho^{\text{env}},{}^{(n-1)}\rho^{\text{act}}]$\\
3) Perform an embedded DFT calculation on the environmental subsystem: 
         ${}^{(n)}v_{emb}^{\text{env}} \rightarrow {}^{(n)}\rho^{\text{env}}$\\
4) Calculate an embedding potential for the active subsystem 
         ${}^{(n)}v_{emb}^{\text{act}}[{}^{(n-1)}\rho^{\text{act}},{}^{(n)}\rho^{\text{env}}]$\\
5) Perform an embedded WFT calculation on the active subsystem: 
         ${}^{(n)}v_{emb}^{\text{act}} \rightarrow {}^{(n)}\rho^{\text{act}}$\\
6) Repeat 2.-5. (increasing $n$ by $1$ each time) until convergence to a given threshold is 
         reached\\
WFT methods accessible in our implementation include Hartree--Fock, CASSCF, DMRG (with and without orbital optimization: DMRG-SCF and DMRG-CI, resp.).
Although the embedding potential for the WFT subsystem depends on the 
density calculated from the wavefunction, we keep it fixed during the wavefunction calculation, which simplifies the 
implementation. The error introduced by this approximation vanishes as soon as the 
wavefunction converges w.r.t.\ the f\&t procedure, thus no additional error is produced in the final results.\\
The implementation provides an interface between a development version of {\sc Molcas}  \cite{molcas7}, 
the {\sc Maquis} quantum chemical DMRG program,\cite{keller2014b} and {\sc Adf} \cite{teVelde:2001a, neug2005e, jaco2008b}. 
The quantum chemical version of {\sc Maquis}  explicitly implements a matrix-product operator based formalism and
exploits all technical features of the underlying {\sc Maquis}\cite{dolfi14} program.
Starting orbitals for the DMRG calculations were taken from a preceding Hartree--Fock calculation with {\sc Molcas}.
We extended the scripting framework {\sc PyADF}\cite{jaco2011} accordingly to allow for easy scripting of the 
calculations. All DFT calculations were performed with {\sc Adf} and the PW91\cite{perd1992c} functional 
in a TZP\cite{vanl2003} basis. Also the embedding potentials were calculated with {\sc Adf} with 
the PW91k\cite{lemb1994} functional for the nonadditive kinetic energy. A supermolecular 
integration grid was used throughout. All DMRG-SCF calculations were performed with {\sc Molcas} using a 
cc-pVTZ\cite{dunn1989} basis. The number of renormalized DMRG block states $M$ was set to 
1000, unless otherwise noted. Structures were optimized with ORCA\cite{nees2012} with 
BP86\cite{beck1988, perd1986}-D3BJ\cite{grim2010}/Def2-TZVP\cite{weig2005} and density fitting.\\ 
As a first technical test, for which we can still obtain CASSCF reference data, we considered the polarization of a methane molecule embedded in the environment of an 
ammonium ion as depicted in Figure \ref{fig:methane_ammonium}. All DMRG-SCF results for an active 
space of eight electrons (the $1s$\ electrons of C are kept frozen) in eight orbitals, CAS(8,8), are 
compiled in Table \ref{tab:methane_in_x}. At equilibrium distance ('eq.'), the effect on the subsystem energy 
is by an order of magnitude more pronounced than at a distance of $10$\ \AA\ between the molecule 
centers. As expected, the polarization of the wavefunction is more pronounced for the calculation on a 
closer distance. The wavefunction-polarization effect is more obvious when investigating the dipole 
moment of the methane molecule as shown in Table \ref{tab:methane_in_x_dipole}. Analogous CASSCF-in-DFT 
calculations (not shown) yield the same results up to the presented accuracy. Furthermore, the 
DMRG-SCF-in-DFT data compares very well with the corresponding DFT-in-DFT results (right column of 
Table \ref{tab:methane_in_x_dipole}).\\
\begin{table}[!]
  \caption{DMRG(8,8)-SCF-in-DFT calculations on methane embedded in an ammoniuim-ion environment at different distances. Energies are given in Hartree. 
  $E^{\text{act}}$ is the energy of the WFT subsystem according to Huang and  
  Carter\cite{huan2011}. In the last column, the energy change due to the polarization of the 
  wavefunction is shown.}
  \footnotesize
  \begin{tabular}{lcc}
  \hline
    \hline
  $\boldsymbol{r}(\text{C--N})$ & $E^{\text{act}}-E^{\text{iso}}$ &   
    $\left(E^{\text{act}}-\int{\rho^{\text{act}}(\boldsymbol{r}) 
    v^{\text{act}}_{\text{emb}}(\boldsymbol{r})d\boldsymbol{r}}\right)-E^{\text{iso}}$ \\ 
  \hline
  10 \AA         & $-5.31$ $\times$ 10$^{-1}$     &
     7.44 $\times$ 10$^{-5}$                                            \\
    5.68 (eq.) \AA & $-9.34$ $\times$ 10$^{-1}$     &
     5.97 $\times$ 10$^{-4}$                                            \\
  \hline
    \hline
  \end{tabular}
  \label{tab:methane_in_x}
\end{table}
\begin{table}[!]
  \caption{Comparison of dipole moments in Debye of methane embedded in an ammonium-ion environment obtained from DMRG(8,8)-SCF-in-DFT and analogous DFT-in-DFT 
    calculations, respectively.}
  \footnotesize
  \begin{tabular}{lcc}
    \hline
    \hline
    $\boldsymbol{r}(\text{C--N})$ & DMRG-in-DFT & DFT-in-DFT \\
    \hline
    10 \AA                & 0.11        & 0.12       \\
    5.68 (eq.) \AA        & 0.33        & 0.36       \\
    \hline
    \hline
  \end{tabular}
  \label{tab:methane_in_x_dipole}
\end{table}
To demonstrate that our new implementation reaches beyond the capabilities of a CASSCF-in-DFT embedding approach,
we studied the dipole moment of a HCN molecule induced by the presence of a second HCN molecule (Figure 
\ref{fig:HCN_dimer}). We selected this HCN dimer as an appropriate benchmark example as it is known that HCN chains show a strong cooperative behaviour in this 
regard.\cite{jaco2014} Table \ref{tab:HCN_dimer_dipole}\ comprises the calculated dipole moments 
obtained using either a supermolecular or a WFT/DFT-in-DFT approach. The reference DMRG-SCF 
calculation on the supermolecular dimer comprised an active space of 20 electrons in 18 orbitals 
(twice the valence complete active space for one HCN molecule, i.e., a CAS(10,9)). 
We observe only small differences in the dipole moments between the DFT and the WFT calculations. A 
qualitatively correct polarization of the embedded HCN molecule compared to its isolated counterpart can be observed. 
An advantage of the subsystem approach is the possibility to investigate the properties of only one subsystem in the presence of 
other subsystems. To demonstrate this feature, we collected the different dipole moments for the 
two different HCN molecules A and B in Table \ref{tab:HCN_dimer_dipole}.
\begin{table}[!]
   \caption{The dipole moment of HCN in the presence or absence of another HCN 
     molecule. Supermolecular results are compared to those from embedding 
     calculations, in which the environment was polarized by a density taken either from a wavefunction or from a DFT 
     calculation, respectively.}
  \footnotesize
  \begin{tabular}{lcccc}
    \hline
    \hline
    Method              & Active   & Polarization of & Dipole moment     \\
                        & fragment & the environment & (active fragment) \\ \hline
    DFT                 & isolated & ---             & 2.95              \\
    DMRG(10,9)-SCF      & isolated & ---             & 3.08              \\ 
    DFT                 & dimer    & ---             & 3.38$^1$          \\
    DMRG$^2$(20,18)-SCF & dimer    & ---             & 3.42$^1$          \\ 
    DMRG(10,9)-SCF      & A        & DMRG            & 3.42              \\
    DMRG(10,9)-SCF      & A        & DFT             & 3.39              \\
    DFT                 & A        & DFT             & 3.39              \\ 
    DMRG(10,9)-SCF      & B        & DMRG            & 3.29              \\
    DMRG(10,9)-SCF      & B        & DFT             & 3.30              \\
    DFT                 & B        & DFT             & 3.30              \\ \hline
    \hline
 \end{tabular}
 \begin{flushleft}
  ${}^1$ Per molecule.\\
  ${}^2$ DMRG parameter $M$ was increased to 2048.
 \end{flushleft}

  \label{tab:HCN_dimer_dipole}
\end{table} 
The importance of taking into account a fully self-consistent polarization of the environment by 
means of a f\&t procedure is indicated in Figure \ref{fig:FaT_convergence}. The f\&t procedure shows 
a fast linear convergence. The error decreases by remarkable $98.6\%$ in each f\&t cycle 
compared to the converged result.
Our new DMRG-in-DFT embedding approach combines the advantages of DMRG with an accurate 
quantum mechanical embedding scheme using a f\&t approach for the polarization of the embedding 
density. This approach facilitates accurate calculations on systems with strong static correlation 
embedded in environments whose effects are important beyond a classical (purely electrostatic) description. 
The pilot applications reported in this communication are proof-of-principle calculations and the algorithms offers
various options for future developments.   
Clearly, perturbative corrections should be considered on top of the converged DMRG-in-DFT results to capture dynamic correlations.
The {\sc Maquis} program is currently extended to produce the required many-body density matrices for different perturbation theories to second order.\\
{\bf Acknowledgements}
TD and JN acknowledge funding by the Deutsche Forschungsgemeinschaft through SFB 858. 
This work was supported by ETH Z\"urich (Research Grant ETH-34 12-2) and by the Schweizerischer Nationalfonds (project 200020\_144458).
Support by COST Action CODECS is gratefully acknowledged.


\end{document}